\documentclass{eas}
\usepackage{graphicx}
\begin{document}

\title{IR imaging surveys of AGB stars in the Magellanic Clouds} 
\author{Maria-Rosa L. Cioni}\address{University of Hertfordshire, STRI, College Lane, Hatfield AL10 9AB, United Kingdom}
\runningtitle{IR images of AGB stars in the Clouds}
\begin{abstract}
AGB stars are ideal IR targets because they are cool and bright.  Most
of them escaped detection in optical or shallow IR surveys in the
eighties contributing to the puzzling missing number of AGB stars with
respect to theoretical predictions and former stages of
evolution. Observations and AGB models have advanced steadily in the
following decades providing us with an almost complete view of the AGB
stars in the Magellanic Clouds. Their properties are tracers of
structure and chemistry across galaxies. New surveys will be able to
fill-in the gaps, in terms of sensitivity and monitoring, providing
new constraints for the formation and evolution of the Magellanic
Clouds.

\end{abstract}
\maketitle
\section{Introduction}
Asymptotic Giant Branch (AGB) stars are post-main sequence stars that
represent the most luminous stage of evolution for low- and
intermediate-mass stars.  One of the main properties of AGB stars is
that they loose $50-80$\% of their mass (gas and dust), they therefore
chemically enrich the interstellar medium (ISM) and represent the main
source of dust in the Universe. AGB stars are also indicators of
distance, structure and metallicity. They contribute to the integrated
light of galaxies and exist in large numbers in the neighbouring
Magellanic Clouds.

All stars with an initial mass comprised between $0.8$ and $8$
M$_\odot$ become AGB stars. The AGB phase is rather brief ($<0.01$
Gyr) compared to other stellar evolutionary phases where stars spend
most of their life, e.g. the main sequence. AGB stars are
characterized by a double shell burning of H and He, respectively,
around their C- and O-rich nucleus. Throughout the AGB structure a
complex chemistry is developed: molecules form right above the
photosphere and these are responsible for two classes of AGB stars,
C-rich (or C-type) where carbonaceous molecules like CN and C$_2$
develop and O-rich (or M-type) where oxides like TiO and VO develop,
after the formation of stable CO. The chemical path-way an AGB star
will take depends on the ISM metallicity prior to their origin and on
the efficiency of the dredge-up process that brings elements
synthesized in the stellar interior to the surface.  At a given
distance from the centre of the star, where temperature and pressure
are appropriate, dust forms, Dust will also be predominantly of
carbonaceous or silicate type and this is related to the molecular
composition in the AGB stars atmospheres. Above the dust layer other
molecules, like OH and HCN, develop in the so-called circumstellar
region.

The AGB phase is a dynamical phase because stars experience surface
luminosity variations with long periods and large amplitudes, and
mass-loss.  These and further details on the AGB stars can be found in
Habing \& Olofsson (\cite{habol}).

\section{Infra-Red images}
Before Infra-Red (IR) images of AGB stars were obtained the Magellanic
Cloud stellar population was characterized by optical
images. Initially the sensitivity was limited to bright super giant
and carbon stars (e.g.~Westerlund et al \cite{wes64}). Later, fainter
sources were studied by Blanco et al (\cite{bl80}) in several fields
in the both the Large Magellanic Cloud (LMC) and the Small Magellanic
Cloud (SMC). These observations were extended to additional fields by
Blanco \& McCarthy (\cite{bl83}). At the same time the variability of
AGB stars was investigated by Hughes \& Wood (\cite{hu90}) who
referred to AGB stars as Long Period Variables (LPVs).

These authors discovered that there are several C-rich AGB
stars. There were, however, not enough luminous AGB stars compared to
the prediction by theory and to the known large number of Cepheid
stars. The latter are the precursors of AGB stars of moderate mass. It
emerged that because AGB stars are cool (red) they are potentially 
obscured by dust preventing their observation in the
optical. Therefore, IR observations were needed to progress in this
field. Note that the deficit of AGB stars was several hundreds!

\subsection{First IR images}
\begin{figure}
\includegraphics[width=12cm]{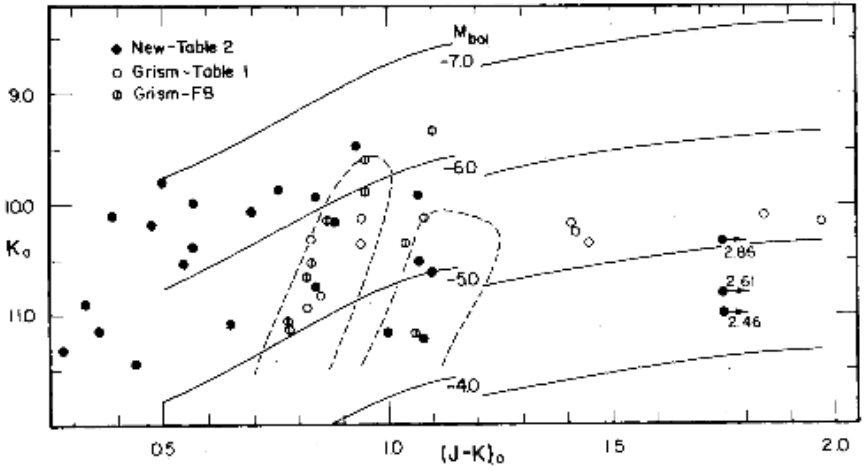}
\caption{Colour-magnitude diagram of LMC bar-west sources (Frogel \&
  Richer \cite{fro83}). Solid lines are lines of constant bolometric
  magnitude while the dashed areas indicate the region occupied by
  upper RGB stars.}
\label{fig1}
\end{figure}

The first IR images of AGB stars in the Magellanic Clouds were
obtained by Frogel \& Richer (\cite{fro83}). They used the CTIO 1.5m
telescope to observe a field in the bar west region of the LMC
(Fig.~\ref{fig1}). The observations were sensitive to stars as faint
as $K=11$ and the spatial resolution corresponded to an aperture of
$27^{\prime\prime}$. A few red, but not very luminous AGB stars, were
identified.

The study of the AGB population advanced in parallel with the
investigation of their variability aspect; this is still the case at
present.  Using the $K$-band magnitude it was discovered that LPVs
obey a period luminosity relation that represents an alternative
distance indicator to Cepheid stars (Hughes et al \cite{hu90}).

\subsection{Mid-IR space observations}
In 1983 the IR Astronomical Satellite (IRAS) scanned the sky. The
instrumentation on board opened a new window into the study of AGB
stars, especially for those surrounded by dust. IRAS also allowed for
making the transition between studying individual AGB stars to using
them in large numbers to investigating properties of the hosting
galaxy (i.e.~the Milky Way).  The exploitation of the IRAS catalogue
is still on-going. IRAS discovered $50$ obscured AGB stars in the LMC
and $25$ in the SMC (van Loon et al \cite{vlo99}). The distribution of
cool stars in the IRAS colour-colour diagram, [12]-[25] versus
[25]-[60], was presented by van der Venn \& Habing
(\cite{vdv89}). Regions occupied by AGB stars of a different type and
with a different dust shell thickness were identified, as well as the
region where Planetary Nebulae (PNe), the successors of AGB stars, are
found. It was suggested that ageing AGB stars develop pulsation
variability and thicker dust shells.

In the following decade the IR Space Observatory (ISO; 1995-1998)
performed targeted observations of the AGB stars discovered by IRAS.
The power of the ISO instrumentation relied both in the imaging and
spectroscopic capabilities that allowed for studying the type of dust,
O-rich if absorption is present at $9.7$ $\mu$m and C-rich if emission is
present at $11.4$ $\mu$m or absorption at $3$ $\mu$m, and to relate this
with the photometric colours and stellar models.

ISO performed also a mini-survey of Magellanic Clouds (Loup et al
\cite{lou99}). This survey covers an area of $0.8$ deg$^2$ and $0.28$
deg$^2$ of the LMC and SMC, respectively. Imaging observations were
obtained in the [$4.5$], [$7$] and [$12$] $\mu$m filters. The
considerably improved spatial resolution and sensitivity of ISO with
respect to IRAS showed that IRAS missed $\sim 50$\% of the dust
obscured AGB stars and that these stars are as luminous as C-rich AGB
stars with thin dust shells (Fig.~\ref{fig3}). The ISO imaging data
were combined with the near-IR data obtained from the DENIS survey
(Cioni et al \cite{cio00a}).

More or less simultaneously mid-IR space observations were obtained
from the Mid-course Space Experiment (MSX; 1996-1997). This satellite
observed $\sim 100$ deg$^2$ of sky at better spatial resolution but
worse sensitivity than IRAS. The colour-colour diagram, [$K$]-[$8$]
versus [$J$]-[$K$], showed obscured AGB stars but with a non-negligible
overlap with PNe and HII regions (Egan et al \cite{ega01}).

\subsection{Near-IR images}

\begin{figure}
\includegraphics[width=6cm]{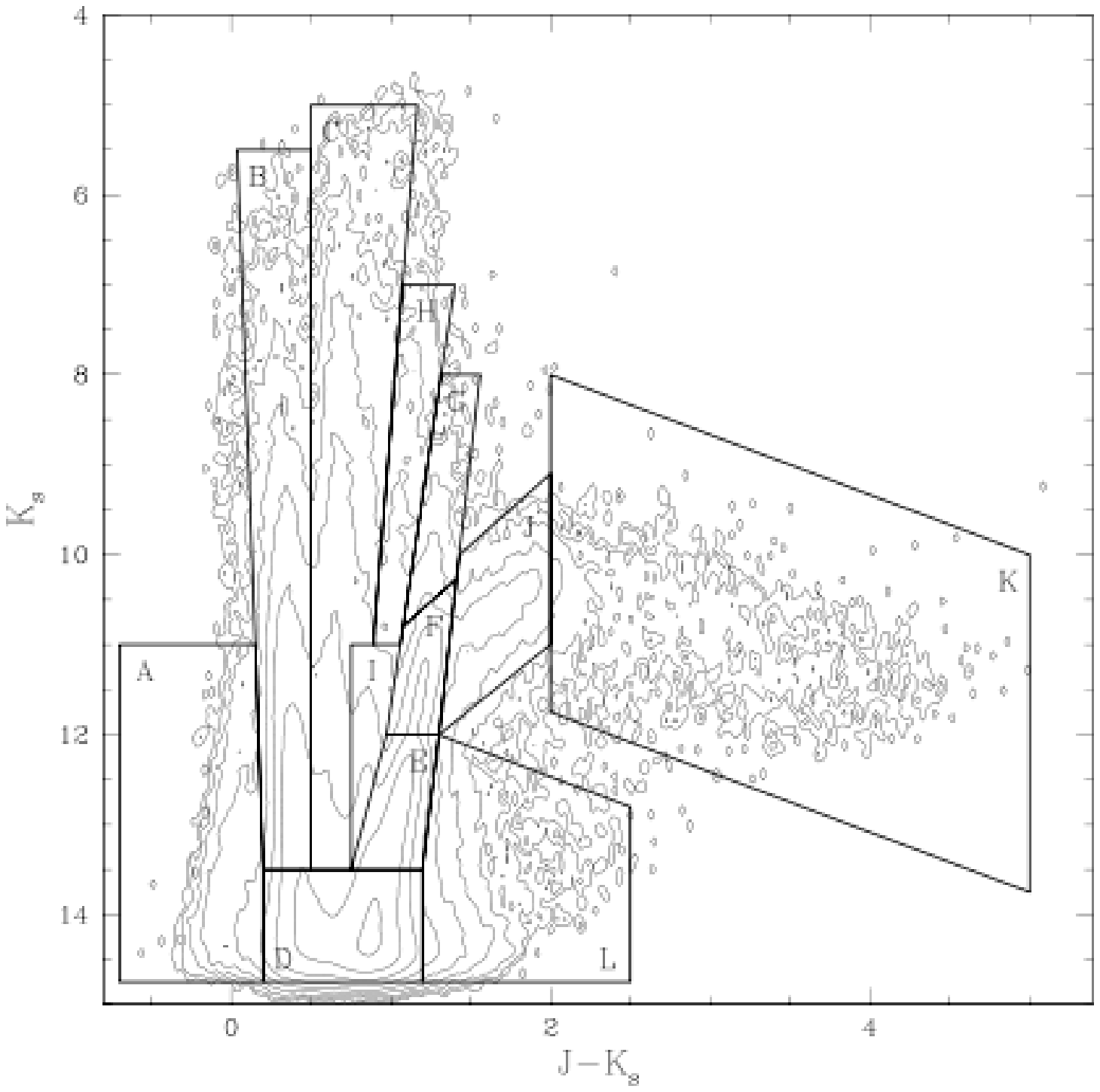}
\qquad
\includegraphics[width=6cm]{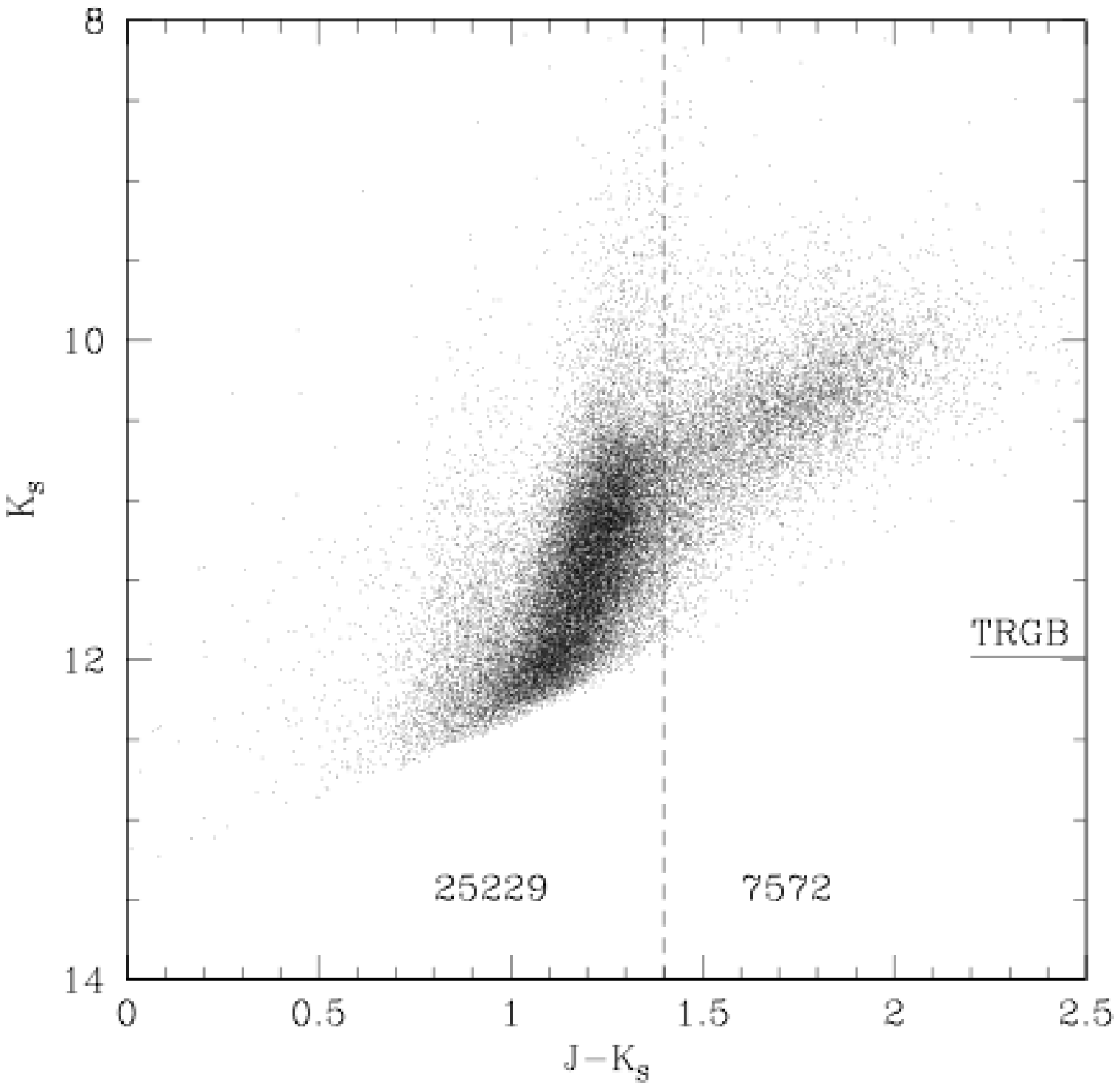}
\caption{(Left) Colour-magnitude diagram of the LMC observed by 2MASS
  where different regions are occupied by different types of stars
  (Nikolaev \& Weinberg \cite{nik00}). (Right) Colour-magnitude
  diagram of AGB stars in the LMC selected from DENIS data, the region
  occupied by O- and C-rich AGB stars is clearly indicated (Cioni \&
  Habing \cite{cio03b}).}
\label{fig2}
\end{figure}

The most comprehensive near-IR sky surveys that provided a wealth of
data for studying the Magellanic Clouds were the 2MASS (1997-2001;
$JHK_s$ filters) and DENIS (1995-2001; $IJK_s$ filters) surveys. Among
their highlights: (i) the distribution of stars in the
colour-magnitude space, e.g.~$J-K_s$ versus $K_s$ (Nikolaev \&
Weinberg \cite{nik00}), and (ii) the distinction between O-rich and
C-rich AGB stars of the earliest spectral types (Cioni \& Habing
\cite{cio03a}) $-$ see Fig.~\ref{fig2}. Both surveys detected the
almost complete population of AGB stars with absent or thin dust
shells, as well as several obscured AGB stars.

The view of the Magellanic Clouds changed dramatically from that of
typical irregular galaxies with central features traced in the optical
by young stars and star forming regions, to smooth and regular
extended structures traced by giant stars (Cioni et al \cite{cio00b})
in the near-IR. The LMC shows a large thick bar confined within an
outer elliptical structure with hints of spiral arms. The SMC
resembles a dwarf elliptical galaxy.  The ratio between C-rich and
O-rich AGB stars, the C/M ratio, provides also a view of the
metallicity distribution within these galaxies (Cioni \& Habing
\cite{cio03a}). This ratio shows a positive gradient within the LMC
and a clumpy distribution consistent with a flat gradient in the SMC.

The conversion and interpretation of the C/M ratio versus [Fe/H]
abundance has been investigated recently by Cioni (\cite{cio09}). The
smoothly declining LMC AGB gradient agrees with that of old (several
Gyr) stellar clusters and RR Lyrae stars, but it does differ
significantly from the gradient traced by young (a few Gyr) red giant
branch (RGB) stars and stellar clusters that appears rather flat. The
latter suggest a flattening of the gradient with time and is perhaps
influenced by the effect of the bar. In the SMC nor AGB stars nor
other indicators such as: RGB stars, PNe, stellar clusters with a
different age, show a significant gradient. This can also be due to
the bar or to the projection effect of two populations with a
different mean age, a young one in the disk and an old one in an outer
spheroid.  Their average metallicity is consistent with that in the
Magellanic Bridge and of the LMC at $\sim 4$ kpc from its centre
supporting tidal stripping resulting from the dynamical interaction
between the Magellanic Clouds.

\subsection{Recent IR surveys}

The Spitzer space telescope, launched in 2003, has surveyed the
Magellanic Clouds in different filters as part of two major projects:
the SAGE and S$^3$MC surveys of the LMC and SMC, respectively. These
surveys have provided a large and homogeneous database for studying
AGB stars, their evolution and mass-loss properties.

\begin{figure}
\includegraphics[width=6cm]{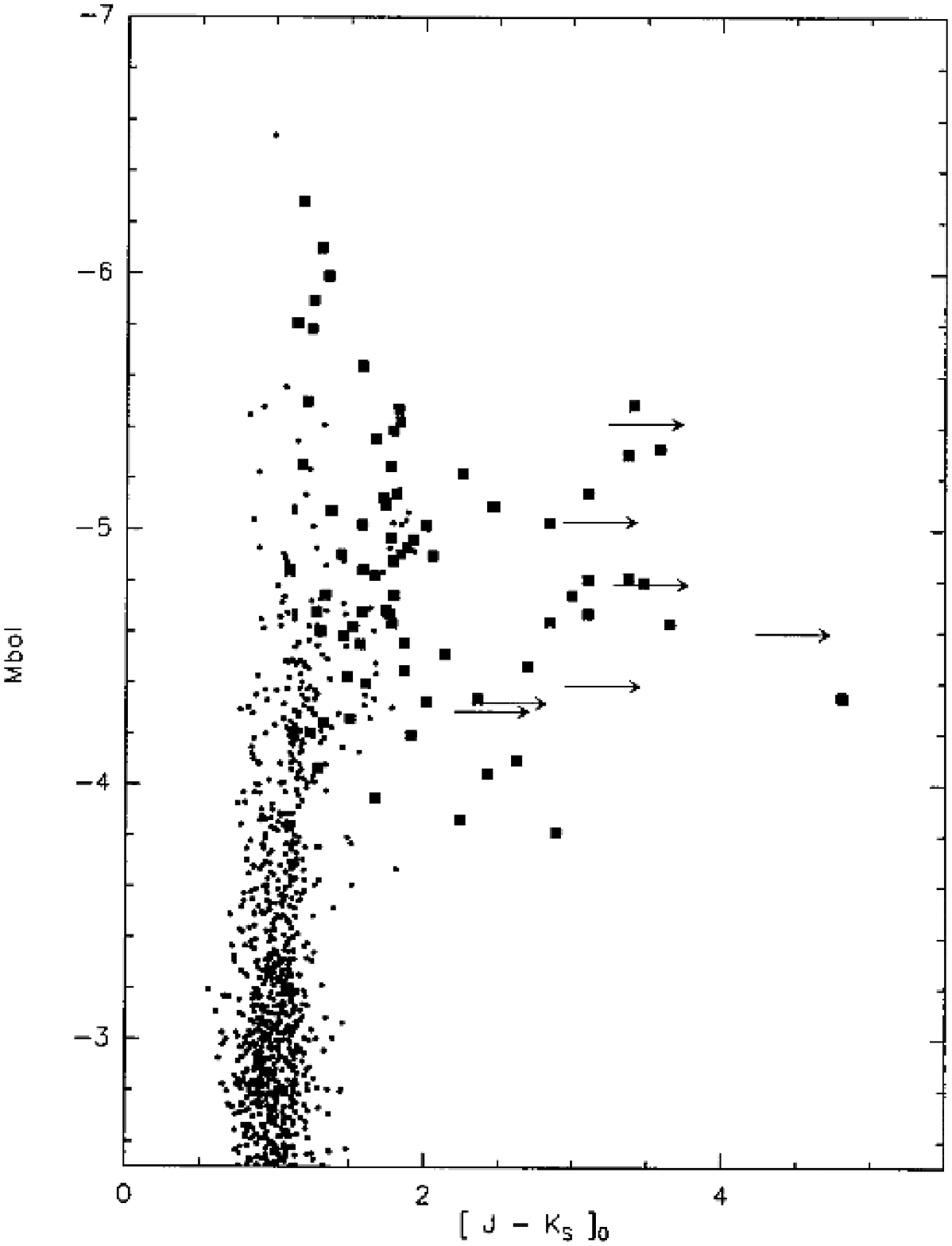}
\qquad
\includegraphics[width=6cm]{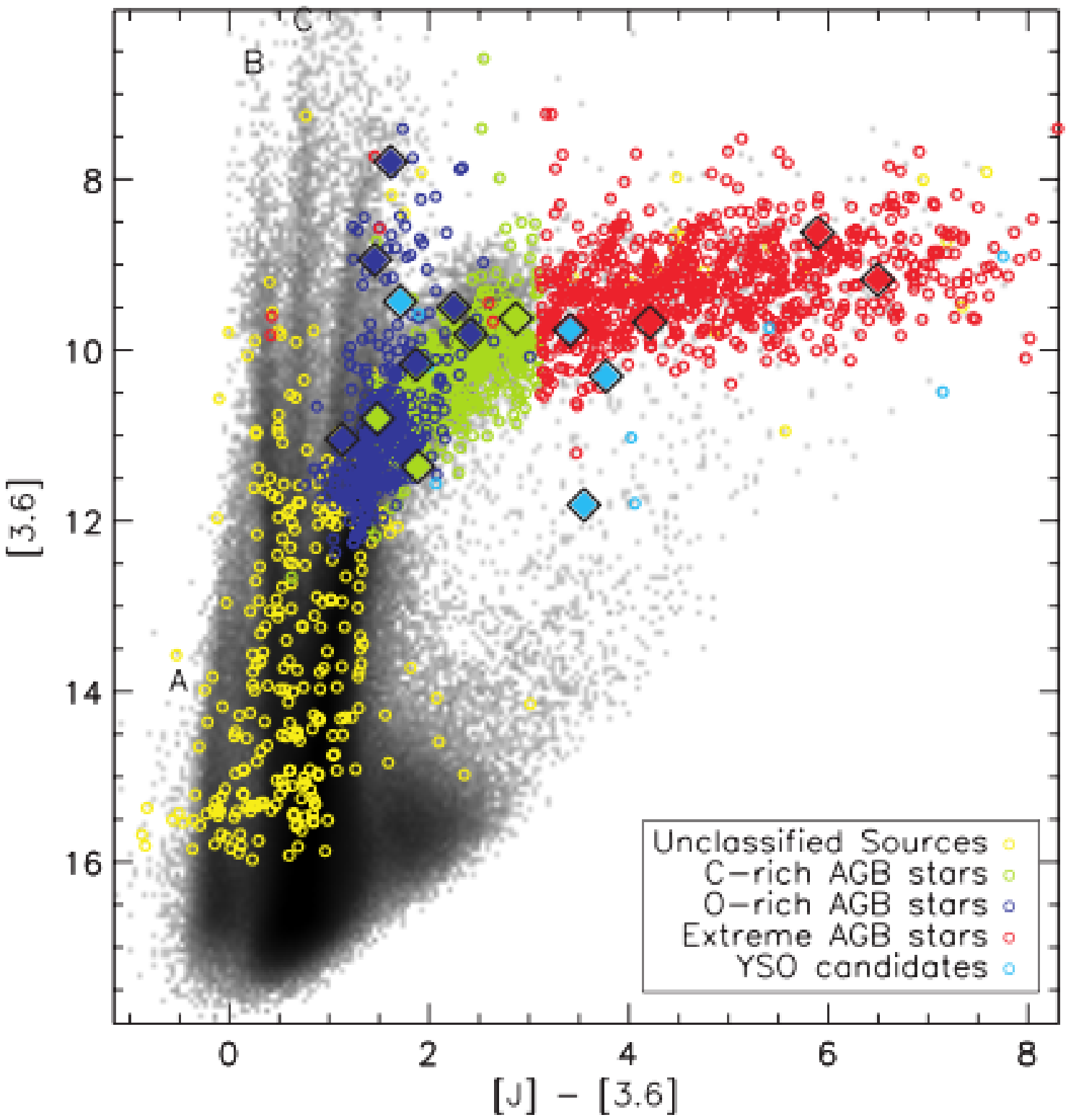}
\caption{(Left) Luminosity distribution of AGB stars from ISO and
  DENIS (Loup \cite{lou99}). Squared symbols indicate ISO matches.
  (Right) Colour-magnitude diagram of AGB stars from Spitzer and 2MASS
  (Vijh et al \cite{vij09}). Coloured symbols indicate Spitzer variables.} 
\label{fig3}
\end{figure}

SAGE (Meixner et al \cite{mei06}) covered $49$ deg$^2$ and acquired two
epochs, separated by three months, that allowed for identifying
variable stars as well as their location in colour-magnitude and
colour-colour diagrams. Vijh et al (\cite{vij09}) discusses the
distribution, variability and dust properties of the SAGE stars:
$66$\% of the extreme/obscured AGB stars are variable, $6.1$\% of the
C-rich AGB stars and $2$\% of the O-rich AGB stars with thin
circumstellar shells are also variables in the mid-IR Spitzer filters
(Fig.~\ref{fig3}).  The spectral energy distribution (SED) obtained
from the combination between 2MASS and Spitzer band-widths shows that
the lack of variability data has a strong influence on the integrated
flux and on the estimated mass-loss rate. The sensitivity and spatial
resolution of Spitzer represent a tremendous improvement versus
previous data. In particular a spatial resolution of
$1-2^{\prime\prime}$ in the $3.5-8.0$ $\mu$m range is comparable to
the resolution of ground based near-IR surveys, consenting a more
secure identification of counterparts. The sensitivity in the mid-IR
has surpassed that in the near-IR leaving unmatched many newly
discovered objects.

The IR survey facility telescope (IRSF; 2001-2006) observed a large
area encompassing the LMC, SMC and the part of the Magellanic Bridge
close to the SMC wing (Kato et al \cite{kat07}). This is the most
sensitive near-IR survey of the Magellanic Clouds to-date reaching a
$10\sigma$ limit at $J=18.8$, $H=17.8$ and $K_s=16.6$. The instrument
resolution corresponds to $0.45$ $^{\prime\prime}$/pix while the
observations were obtained with an average seeing of
$1.2^{\prime\prime}$.  AGB stars were not easily found in the Bridge
and in general no AGB stars were found down to $K_s=13.5$ at
$J-K_s=6$.

The AKARI telescope, launched in 2006, has completed the observation
of $10$ deg$^2$ in the north-east area of the LMC and is currently
undertaking an all-sky survey covering the range $1.8-180$ $\mu$m. A
preliminary catalogue of point sources has been published by Ita et al
(\cite{ita08}). The band-widths are similar to Spitzer, but for the
$11\mu$m one, and have been matched with IRSF near-IR magnitudes.
Their combination reaches sources $\sim 0.5$ mag fainter than the
Spitzer-2MASS combination. The colour-magnitude diagram, [$3$]-[$11$]
  versus [$11$], shows interesting new features traced by AGB stars: a
  faint plume of sources just brighter than the RGB tip and bending to
  red colours indicating O-rich giants with Al oxide dust (Blum et al
  \cite{blu06}, Lebzelter et al \cite{leb06}).

\subsection{Forthcoming IR surveys}

In the near-IR domain the VISTA wide-field telescope is currently
being commissioned at the European Southern Observatory (ESO).  The core
programme for the next five years at VISTA includes six public surveys
of which two are devoted to the observation of stars, the others are
extragalactic surveys, and one is focused on the Magellanic system.

\begin{figure}
\includegraphics[width=12cm]{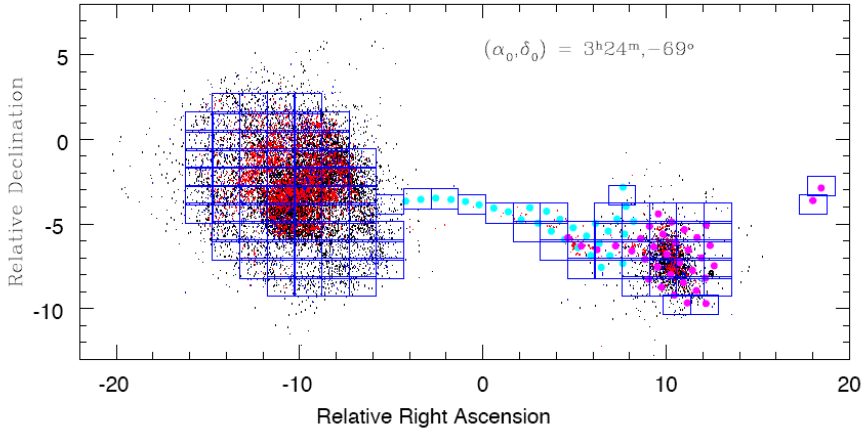}
\caption{Distribution of VISTA tiles across the Magellanic
  System. Underlying small dots indicate the distribution of C stars,
  clusters and associations while thick dots show the location of
  observations to be performed with the VST in the optical.}
\label{fig4}
\end{figure}

The VISTA survey of the Magellanic system (VMC; Cioni et al
\cite{cio08}) will provide the missing link to optical surveys, with a
similar sensitivity, as well as counterparts for the mid-IR
sources. VMC will cover $180$ deg$^2$ distributed across broad areas
in the LMC and SMC, the entire length of the Magellanic Bridge
connecting the two galaxies (for a width of $1.5-4.5$ deg) and a
couple of fields in the Magellanic Stream (Fig.~\ref{fig4}).

Observations will be obtained in three filters, $YJK_s$ providing a
$10\sigma$ limit at $Y=21.9$, $J=21.4$ and $K_s=20.3$. They will be
executed in service mode to guarantee homogeneous conditions and data
quality. The instrument resolution is $0.51$ $^{\prime\prime}$/pix and
observations will be obtained with an average seeing of
$0.6-0.8^{\prime\prime}$ depending on crowding.  VMC is also a
multi-epoch survey because it will reach its nominal sensitivity by
combining $12$ independent epochs in $K_s$, and $3$ in $Y$ and $J$,
respectively. The completion of the survey requires $1840$ hours.

The main science goals of the VMC surveys are to derive the spatially
resolved star formation history (SFH) and to measure the
three-dimensional (3D) structure of the Magellanic system. The VMC
data will also be used to find stellar sub-structures (clusters and
streams), emission line objects (like PNe), to derive distances and
measure proper motions, to study star formation, to re-construct the
system using dynamical models and eventually find extra-galactic
objects (star-burst galaxies and dusty active galactic nuclei at high
redshift), as well as form many other scientific applications.

The VMC survey will be complementary to an on-going optical survey of
the outer regions of the Magellanic Clouds ($8-20$ kpc from the galaxy
centres). This area will also be surveyed by VISTA as part of the
VISTA Hemisphere Survey (VHS) but to a shallower depth than the VMC
depth, and to the space astrometry mission GAIA that will measure the
metallicity and motion of evolved giant stars of the Magellanic
System.

Simulations of the stellar population that VMC will detect show that
stars just below the oldest main-sequence turn-off will be well
detected. The increased in the parameter space, e.g.~near-IR
photometry, at this depth will allow for determining the metallicity
and age distribution with improved accuracy (Kerber et al
\cite{ker09}). The 3D structure will be measured using different
distance indicators and in particular the period-luminosity relation
for short period variable stars, RR Lyrae stars and Cepheids. VMC will
provide the near-IR magnitude and the period will be obtained from
large optical catalogues in the literature (e.g. OGLE-III and EROS-II)
or from observations at the VLT Survey Telescope (VST), for the Bridge
and SMC parts only.

Although VMC is a deep survey targeting faint stellar populations it
will find counterparts to faint AGB stars and, via an accurate
analysis of the SFH of the system, it will relate them to their
progenitors.

\subsection{Future surveys}

The Magellanic Clouds will be targeted by different new
missions. During this meeting the Herschel satellite was successfully
launched. Herschel will probe the mid- to far-IR regime for AGB stars
in the Magellanic Clouds.

The next step in ground-based IR astronomy may be taken really in
Antarctica with the development of a telescope like PILOT (see other
contributions in this proceeding). This will be a 2m class telescope
that will explore the dark side of the $K$-band and possibly extend to
the $L$-band, reaching an unprecedented depth, and will also be ideal
to monitor AGB stars in the Magellanic Clouds.

\section{Conclusions}

IR imaging of AGB stars in the Magellanic Clouds were aimed at finding
the most luminous and dusty, thus obscured, sources. Different
criteria were developed to distinguish and classify AGB stars into O-
and C-rich. Multi-wavelength and multi-epoch (optical only)
observations have allowed to quantify mass-loss rates and the coupling
with AGB pulsation and evolution. 

Major progress in the study of AGB stars has occurred not only on the
observational side but also on the theoretical side with the
development of models that are able to interpret the location of AGB
stars in the IR (Marigo et al \cite{mar08}). These authors have
produced isochrones including molecular opacities in O- and C-rich AGB
stars, the hot bottom burning process, the effect of the pulsation
mode (first overtone or fundamental mode) to the AGB lifetime and the
mass-loss with respect to the different surface chemistry. By
exploring the range of ages typical for AGB stars, these models,
reproduce well the distribution of the AGB population in the
Magellanic Clouds in both near- and mid-IR diagrams.

The most luminous AGB stars experience the strongest mass-loss
rate. This result has been greatly re-affirmed with the investigation
by Fraser (\cite{fra08}) that involves the combination between Spitzer
and 2MASS data with optical monitoring data from the MACHO
project. The bolometric magnitude and the mass-loss rate were obtained
from the SED of individual AGB stars and the period extracted from
their optical light-curve. The major uncertainty still present in the
period-luminosity relations is directly reflected in the uncertainty
attributed to the bolometric magnitude. To obtain average bolometric
magnitudes it is necessary to monitor the SED across a few years. The
reduced scatter around the period-luminosity relations will also
contribute to refining AGB stars as a powerful tool to measure
distances in the Universe.  

There is at present a paucity of IR monitoring ($2$ epochs by Spitzer,
$12$ epochs by VMC) and a lack of foreseen projects to improve this
condition.  Mean bolometric luminosities are the next step in the
investigation of the mass-loss mechanism and in measuring distances
from the period luminosity relation.  The Magellanic Clouds are the
closest examples of interacting galaxies and represent an ideal
laboratory for stellar evolution. The design of PILOT, a telescope for
the Antarctica site, is well suited to advance the study of AGB stars
in these neighbouring galaxies by providing better sensitivity,
spatial accuracy and wide-field monitoring.


\end{document}